\documentclass[aps,amsmath,showpacs,amsfonts,10pt]{revtex4}
\usepackage{epsfig,graphicx}
\usepackage[english]{babel}
\usepackage{amsfonts}
\usepackage{amsmath}
\usepackage{latexsym}
\usepackage{graphics,bm}
\usepackage{dcolumn}
\usepackage{bm}
\usepackage{rotating}

\begin{document}

\title{Optomechanical Entanglement between an Ion and an Optical Cavity Field}

\author{Aranya B Bhattacherjee}
\affiliation{School of Physical Sciences, Jawaharlal Nehru University, New Delhi-110067, India }

\begin{abstract}
We study an optomechanical system in which the mechanical motion of a single trapped ion is coupled to a cavity field for the realization of a strongly quantum correlated two-mode system. We show that for large pump intensities the steady state photon number exhibits bistable behaviour. We further analyze the occurrence of normal mode splitting (NMS) due to mixing of the fluctuations of the cavity field and the fluctuations of the ion motion which indicates a coherent energy exchange. We also find that in the parameter regime where NMS exists, the steady state of the system shows continuous variable entanglement. Such a two-mode optomechanical system can be used for the realization of continuous variable quantum information interfaces and networks.
\end{abstract}


\maketitle

\section{Introduction}

Cold trapped ions represent elementary quantum systems that can be isolated from the environment. These trapped ions can be brought nearly to rest by laser cooling and both their internal electronic degree of freedom and external motion can be coupled and manipulated by external lasers. This unique property makes the cold trapped ions ideally suited for quantum optical studies under well controlled conditions.

Cold trapped ions are one of the most promising physical systems for implementing quantum computation \citep{1} and quantum simulations \citep{2}, search for new physics beyond the Standard Model \citep{3}, ultra-high precision measurements \citep{4} and extremely accurate frequency standards \citep{5}. The long coherence times and individual addressing of ions makes them suitable for the experimental implementation of quantum gates and quantum computing systems. Recent experimental progress with cold trapped ions include simulating a quantum magnet with two-trapped ions \citep{6}, demonstration of a phonon laser \citep{7} and realization of a quantum phase transition of polaritonic excitations using two trapped ions \citep{8}. The motion of the center-of-mass (CM) of an ion confined in a Paul trap can be regarded as a quantum harmonic oscillator. The interaction between the CM and internal degrees of freedom of the ion can be coupled by an external laser light in the Jaynes-Cummings (JC) model \citep{9}. Based on this JC model, a fundamental logic gate has already been demonstrated experimentally \citep{10}, together with nonclassical motion of states of a single ion \citep{11}.

In the present work, we study an optomechanical system comprising of a single trapped ion coupled to a single optical cavity mode. We study the system in a highly detuned limit so that we can eliminate the electronic degree of freedom of the ion. We show that coupling of the fluctuations of the mechanical motion of the ion and the fluctuation of the cavity field leads to the splitting of the normal mode into two modes (Normal Mode Splitting). The normal mode splitting (NMS) demonstrates a coherent energy exchange between the mechanical and the optical mode. It is shown that motional and field fluctuations are entangled in the parameter regime where NMS is found. Thus, in this hybrid system, quantum information exchange takes place between the ion and the cavity field. Earlier studies related to phonon-photon interactions with a trapped ion in a cavity was reported by Massoni et. al. \citep{12} and Orsag \citep{13}. They showed the squeezing transfer from vibrations to a cavity field. More recently, It was shown theoretically, entanglement between motional mode of an ion chain and a cavity field exists where mechanical bistabiltty is found \citep{14}. During the past several years, several theoretical papers on ion-cavity QED context have discussed issues related to entanglement generation, preparation of nonclassical states, or realization of quantum gates \citep{pel,enk,plen,zhen,pach,loug,chim,li,li2,chim2,bina,hark}

Entanglement is one of the important elements of quantum mechanics as it is responsible for correlations between observables \citep{bel}. It is now rigorously investigated in performing quantum communication and quantum computation with a high degree of efficiency \citep{nielsen}. A useful and detailed review of the theory of entanglement in systems of continuous variables was recently given by Eisert and Plenio in \citep{eisert1}. Some of the interesting experimental demonstrations in this direction are given by the entanglement between collective spins of atomic ensembles \citep{jul}, and between Josephson-junction qubits \citep{ber}.

The simplest scheme capable of generating stationary optomechanical entanglement was studied with a single movable Fabry-Perot cavity mirror \citep{vitali1}, or both movable cavity mirrors \citep{vitali}. In fact, entangled optomechanical systems have exciting application in realizing quantum communication networks, in which the mechanical modes play the role of local nodes where quantum information can be stored and retrieved, and optical modes carry this information between the nodes. This type of scheme is proposed based on free-space light modes scattered by a single reflecting mirror in Refs. \citep{man1,pir,pir1,pir2} . This enables the implementation of continuous variable (CV) quantum teleportation \citep{pir2}, quantum telecloning \citep{pir}, and entanglement swapping \citep{pir1}. Further, entanglement in the steady state of a system is significant because it is stationary, i.e., it has a virtually infinite lifetime, and hence could be used repeatedly.

\section{The Model}

We consider a single two-level trapped ion placed inside an optical cavity. We assume that the trapped ion satisfies the Lamb-Dicke limit. The optical cavity supports a single mode quantized radiation field with frequency $\omega_{c}$. We assume for simplicity that the motion of the ion is restricted to the $x-$ direction i.e along the standing wave traveling in the $x-$ direction. The trapped ion experiences a dipolar transition, formed by the electronic ground state $|g>$ and the excited state $|e>$ at the ionic transition frequency $\omega_{a}$, coupling strongly with the single mode cavity frequency $\omega_{c}$. The cavity is pumped by an external laser at frequency $\omega_{p}$. The dynamics of the system comprising of the single cavity mode and the ion is described by the Hamiltonian,

\begin{equation}
H=H_{cav}+H_{el}+H_{ion}+H_{int},
\end{equation}

where $H_{cav}=-\hbar \Delta_{c} a^{\dagger}a-i \hbar (\eta^{*} a-\eta a^{\dagger})$ is the Hamiltonian for the cavity mode in a reference frame rotating with the pump frequency $\omega_{p}$. Here $a^{\dagger}$ and $a$ are the creation and annihilation operators of a cavity photon, $\Delta_{c}=\omega_{p}-\omega_{c}$ is the detuning between the laser and cavity-mode frequency, while the frequency $\eta$ denotes the pump strength. The term $H_{el}=-\hbar \Delta_{a} \sigma^{+} \sigma^{-}$ is the internal ionic dynamics, with $\Delta_{a}=\omega_{p}-\omega_{a}$ the detuning of the pump from the electronic transition and $\sigma^{+}$, $\sigma^{-}$ are the usual pseudo-spin operators, $\sigma^{+}=|e><e|$, $\sigma^{-}=|g><e|$. The term $H_{ion}=\hbar \omega_{m} b^{\dagger}b$ accounts for the vibrational energy of the ion with $\omega_{m}$ as the frequency of the center-of-mass (CM) vibration and $b(b^{\dagger})$ denoting the destruction and creation operators of the CM vibrational motion mode. Finally, the dipolar coupling between the cavity and the ion reads in the rotating wave approximation (RWA), $H_{int}=\hbar g(x) (\sigma^{+} a + \sigma^{-} a^{\dagger})$, where $g(x)=\frac{\Omega \sin(\tilde{\eta} \hat{x}+\phi)}{2}$. Here $\Omega$ is the coupling parameter directly proportional to the ion-radiation interaction strength (i.e the electric dipole moment of the ion). Here $\hat{x}=b^{\dagger}+b$ denotes the dimensionless position operator for the CM of the ion. Also $\tilde{\eta}=k \sqrt{\hbar/2m\omega_{m}}$ is the Lamb-Dicke parameter with $k$ as the wavenumber, $m$ is the mass of the ion. $\phi$ accounts for the relative position of the CM of the ion to the standing wave. When the ion is centered at the node of the standing wave, $\phi=0$.

We will work in the large ionic detuning limit  (i.e large $\Delta_{a}$) so that we can adiabatically eliminate the ionic degree of freedom by putting $\dot{\sigma^{-}}=\frac{i}{\hbar}[H,\sigma^{-}]$. This finally leads to the optomechanical Hamiltonian,

\begin{equation}
H_{om}=\hbar \Delta a^{\dagger}a+\hbar \omega_{m} b^{\dagger}b+\hbar g_{o}(b^{\dagger}+b)+\hbar g_{o}a^{\dagger}a(b^{\dagger}+b)-i\hbar (\eta^{*}a-\eta a^{\dagger}),
\end{equation}

where, $\Delta=(\frac{\Omega^{2}}{4 \Delta_{a}}-\Delta_{c})$, $g_{o}=\frac{\Omega^{2}\tilde{\eta}\sin{2 \phi}}{4 \Delta_{a}}$.

\begin{figure}[h]
\hspace{-0.0cm}
\includegraphics [scale=0.70] {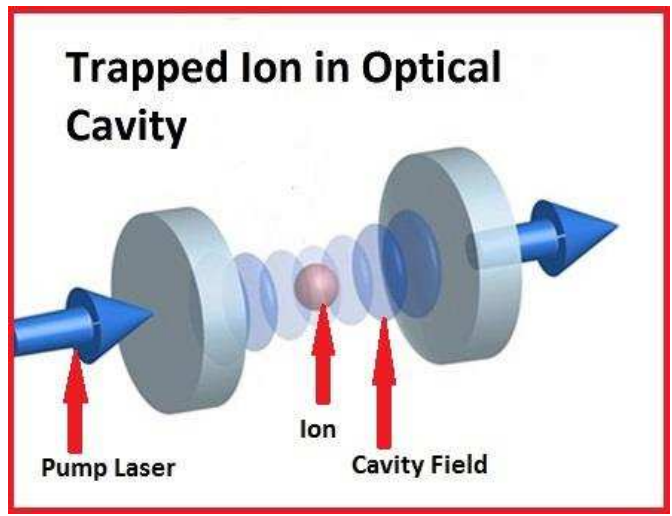}
\caption{Schematic diagram of a single trapped ion in an optically pumped cavity. The mechanical mode of the ion is coupled the optical mode.}
\label{f1}
\end{figure}

\section{Optomechanical Bistability}

One of the hallmark of cavity optomechanical system is the optical bistability via optical pumping, which has been observed in various systems experimentally. In the current system, it is the nonlinearity due to the coupling between the cavity mode and the CM motion of the ion that is responsible for bistable behavior. In order to demonstrate the bistable behavior, we start with the Heisenberg equation of motion for the cavity mode and the mechanical mode of the ion,

\begin{equation}
\dot{a}=-i \Delta a-i g_{o}(b^{\dagger}+b)a-\gamma a-\eta,
\end{equation}

\begin{equation}
\dot{b}=-i \omega_{m}b-i g_{o}a^{\dagger}a-ig_{o}-\Gamma b,
\end{equation}

where $\gamma$ and $\Gamma$ is the cavity damping and the mechanical damping respectively. Substituting the steady state value of the mechanical mode $b_{s}$ into the steady state expression for the cavity mode $a_{s}$ leads to the cubic equation for $a_{s}$, which shows typical bistable behavior.

\begin{equation}
G^{2}x_{m,s}^{3}+G 2(\Delta+G)x_{m,s}^{2}+(\gamma^{2}+(\Delta+G)^{2})x_{m,s}=|\eta|^{2},
\end{equation}

where, $x_{m,s}=|a_{s}|^{2}$ and $G=\frac{2 g_{0}^{2} \omega_{m}}{(\Gamma^{2}+\omega_{m}^{2})}$. Clearly it is seen that the bistability in the intracavity photon number disappears when we have $\phi=n \pi$ and $\phi=n \pi/2$.

\begin{figure}[h]
\hspace{-0.0cm}
\includegraphics [scale=0.70] {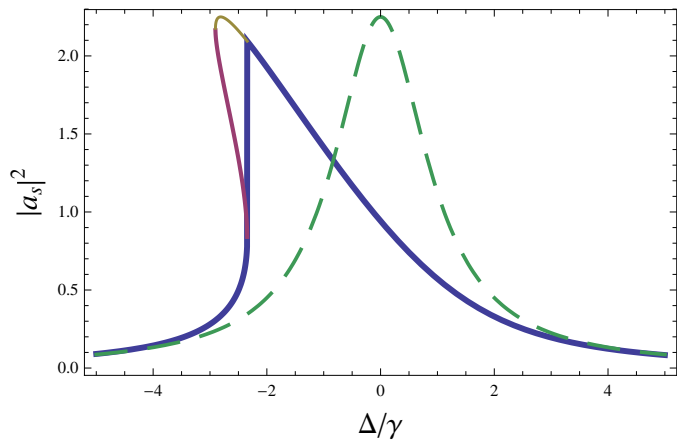}
\caption{Plot of steady state intra-cavity photon number $|a_{s}|^{2}$ as a function of $\Delta/\gamma$ for $\phi=\pi/4$ (solid lines) and $\phi=0$(dashed line). The parameters taken are $\eta/\gamma=150$, $\tilde{\eta}=1$, $\Gamma/\gamma=0.01$, $\Delta_{a}/\gamma=-10$, $\omega_{m}/\gamma=10$.}
\label{f2}
\end{figure}

In Fig.2, we represent $|a_{s}|^{2}$ as a function of $\Delta/\gamma$ for $\phi=\pi/4$ (solid lines) and $\phi=\pi/2$(dashed line).The system shows bistability only for higher pump intensity and $\phi=\pi/4$. Bistability in the intracavity photon number disappears when we have $\phi=0$.
For pump rates higher than a certain critical value, we find three steady state solutions for the mirror displacement, with two of them being stable.The system prepared below resonance will follow the steady state branch until reaching the lower turning point, where a non-steady-state dynamics is excited. This dynamics is governed by the time scale of the mechanical motion of the ion because the cavity damping is almost two orders of magnitude faster ($\Gamma<<\kappa$).

\section{Fluctuations:Normal Mode Splitting}

Here we show that the coupling of the fluctuations of the ion motion and the cavity field fluctuations leads to the splitting of the normal mode into two modes (Normal Mode Splitting(NMS)). The optomechanical NMS however involves driving two parametrically coupled nondegenerate modes out of equilibrium. The NMS does not appear in the steady state spectra but rather manifests itself in the fluctuation spectra of the ion displacement. We now derive the Heisenberg-Langevin equations for the canonical variables and introduce the input noise operators $a_{in}(t)$ and $b_{in}(t)$. We shift the canonical variables to their steady state values, $a\rightarrow a_{s}+a$ and $b\rightarrow b_{s}+b$ and linearize to obtain the following Heisenberg-Langevin equations:

\begin{equation}
\dot{a}=(-i \tilde{\Delta}-\gamma)a-ig_{2}(b^{\dagger}+b)+\sqrt{2 \gamma} a_{in},
\end{equation}

\begin{equation}
\dot{b}=(-i \omega_{m}-\Gamma)a-ig_{2}(a^{\dagger}+a)+\sqrt{2 \Gamma} b_{in},
\end{equation}

where $\tilde{\Delta}=\Delta+g_{1}$, $g_{1}=g_{o}(b_{s}+b_{s}^{*})$, $g_{2}=g_{o}|a_{s}|$. We will assume  throughout our work that $\Gamma<\gamma$. The input noise operators for the optical field satisfies: $<a_{in}(t)>=0$, $<a_{in}^{\dagger}(t')a_{in}(t)>=n_{a}\delta(t'-t)$, and $<a_{in}(t')a_{in}^{\dagger}(t)>=(n_{a}+1)\delta(t'-t)$. The noise operator $b_{in}(t)$ is considered to be classical thermal noise input for the mechanical degree of freedom of the ion which satisfies: $<b_{in}(t)>=0$, $<b_{in}^{\dagger}(t')b_{in}(t)>= <b_{in}(t')b_{in}^{\dagger}(t)>=n_{b}\delta(t'-t)$. The quantities $n_{a}$ and $n_{b}$ are the equilibrium occupation numbers for the optical and mechanical degrees of freedom respectively.
The equations (6) and (7) and their Hermitian conjugates constitutes a system of four first order coupled equations. The stability of the system is given by the Routh-Hurwitz criterion which is $2 g_{2}<\sqrt{(\tilde{\Delta}^{2}+\gamma^{2})\omega_{m}/|\tilde{\Delta}|}$ $\approx$ $\omega_{m}$ for $\omega_{m}>>\gamma$ and $|\tilde{\Delta}|\approx\omega_{m}$.
The steady state displacement spectrum $S_{b}(\omega)$ for the mechanical quadrature $x/x_{o}=b+b^{\dagger}$, $x_{o}=\sqrt{\frac{\hbar}{2m_{eff} \omega_{m}}}$ is given by,

\begin{equation}
S_{b}(\omega)=\frac{x_{o}^2 \omega_{m}^{2}}{2 \pi}|\chi(\omega)|^{2}\left( 2 \Gamma n_{b}+\frac{(\tilde{\Delta}^{2}+\omega^{2}+\gamma^{2})}{2 \tilde{\Delta}\omega_{m} }\Gamma_{b}(\omega)\right),
\end{equation}

\begin{equation}
\chi^{-1}(\omega)= \omega_{m}^{2}+2 \omega_{m} \Omega_{b}(\omega)-\omega^{2}-i \omega(2 \Gamma+\Gamma_{b}(\omega)),
\end{equation}

\begin{equation}
\Omega_{b}(\omega)= g_{2}^{2} \left( \frac{\omega-\tilde{\Delta}}{(\omega-\tilde{\Delta})^2+\gamma^{2}}-\frac{\omega+\tilde{\Delta}}{(\omega+\tilde{\Delta})^2+\gamma^{2}}\right),
\end{equation}

\begin{equation}
\Gamma_{b}(\omega)= \frac{g_{2}^{2}}{\omega} \left( \frac{2 \omega_{m} \gamma}{(\omega-\tilde{\Delta})^2+\gamma^{2}}-\frac{2 \omega_{m} . \gamma}{(\omega+\tilde{\Delta})^2+\gamma^{2}}\right),
\end{equation}
Here $\chi(\omega)$ is the mechanical susceptibility of the ion motion.

\begin{figure}[h]
\hspace{-0.0cm}
\includegraphics [scale=0.70] {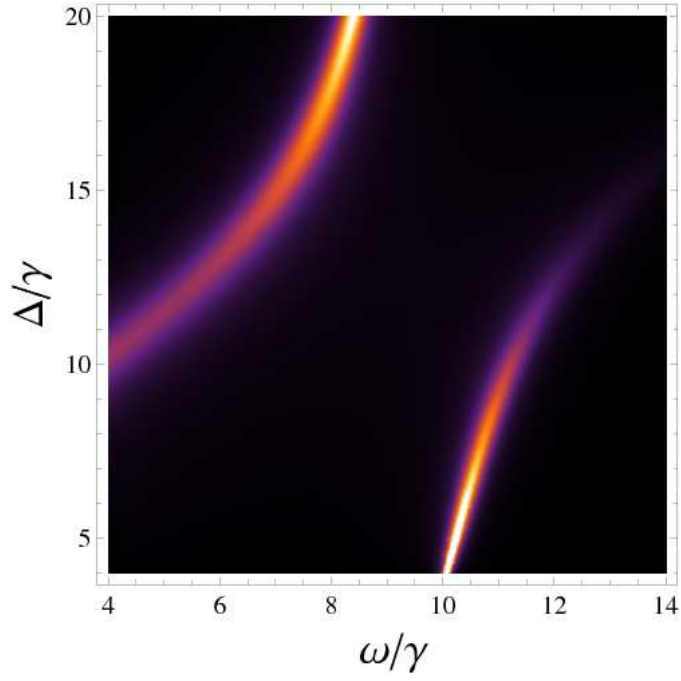}
\caption{Plot of the displacement spectrum $S_{b}(\omega)$ of the ion versus normalized frequency and the normalized effective detuning ($\Delta/\gamma$).The parameters taken are  $\omega_{m}/\gamma=10$, $\Gamma/\gamma=0.01$, $g_{0}/\gamma=-3.0$,$n_{b}=100$.}
\label{f3}
\end{figure}

Figure 3 shows the plot of the displacement spectrum $S_{b}(\omega)$ of the ion versus normalized frequency and the normalized effective detuning ($\Delta/\gamma$). We clearly observe the usual normal mode splitting (NMS). The NMS is associated with the mixing between the mechanical mode of the ion and the fluctuation of the cavity field around their steady states.The origin of the fluctuations of the cavity field is the beat of the pump photons with the photons scattered from the trapped ion. The mechanical mode and the optical mode forms a system of two coupled oscillators exchanging energy. An important point to note is that in order to observe the NMS, the energy exchange between the two modes should take place on a time scale faster than the decoherence of each mode. On the negative detuning side, we did not observe any NMS due to the onset of parametric instability. NMS implies cooling of the mechanical mode.

\section{Continuous Variable Entanglement}

In order to show that the steady state of the system shows continuous variable (CV) entanglement between the cavity field and the mechanical mode of the trapped ion in an experimentally accessible parameter regime, we begin with the equations of motion for the position and momentum quadrature of the optical and the mechanical mode of the system.

\begin{figure}[h]
\hspace{-0.0cm}
\includegraphics [scale=0.70]{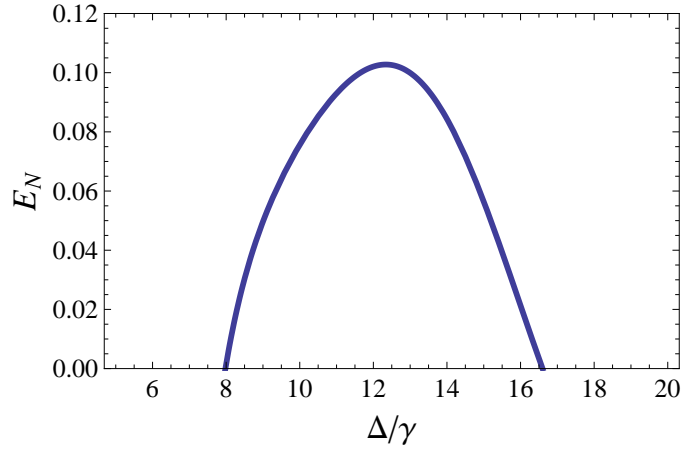}
\caption{Plot of logarithmic negativity $E_{N}$ of the coupled ion-cavity system versus the normalized detuning ($\Delta/\gamma$) for $n_{b}=10$. We have taken $\Gamma=0.01\gamma$, $\omega_{m}=10 \gamma$, $g_{0}=-3.0 \gamma$.}
\label{f4}
\end{figure}

\begin{equation}
\dot{X_{a}}=\tilde{\Delta}P_{a}-\gamma X_{a}+\sqrt{2 \gamma} X_{a}^{in},
\end{equation}

\begin{equation}
\dot{P_{a}}=-\tilde{\Delta}X_{a}-\gamma P_{a}-2 g_{2}X_{b}+\sqrt{2 \gamma} P_{a}^{in},
\end{equation}

\begin{equation}
\dot{X_{b}}=\omega_{m} P_{b}-\Gamma X_{b}+\sqrt{2 \Gamma} X_{b}^{in},
\end{equation}

\begin{equation}
\dot{P_{b}}=-\omega_{m} X_{b}-\Gamma P_{b}-2 g_{2}X_{a}+\sqrt{2 \Gamma} P_{b}^{in},
\end{equation}

where, $X_{a}=\frac{(a^{\dagger}+a)}{\sqrt{2}}$, $P_{a}=\frac{i(a^{\dagger}-a)}{\sqrt{2}}$, $X_{b}=\frac{(b^{\dagger}+b)}{\sqrt{2}}$, $P_{b}=\frac{i(b^{\dagger}-b)}{\sqrt{2}}$, $X_{a}^{in}=\frac{(a^{\dagger}_{in}+a_{in})}{\sqrt{2}}$, $P_{a}^{in}=\frac{i(a^{\dagger}_{in}-a_{in})}{\sqrt{2}}$, $X_{b}^{in}=\frac{(b^{\dagger}_{in}+b_{in})}{\sqrt{2}}$, $P_{b}^{in}=\frac{i(b^{\dagger}_{in}-b_{in})}{\sqrt{2}}$.

The system of above linearized equations of motion can be written in the following compact form,

\begin{equation}
\dot{R(t)}=M R(t)+N(t),
\end{equation}

where, $R(t)=({X_{b},P_{b},X_{a},P_{a}})^{T}$ and the noise vector $N(t)=({\sqrt{2 \Gamma}X_{b}^{in}, \sqrt{2 \Gamma}P_{b}^{in}, \sqrt{2 \gamma} X_{a}^{in}, \sqrt{2 \gamma} P_{a}^{in}})$. Here $M$ is the drift matrix, given as:

\begin{equation}
 M = \left( \begin{array}{cccc}
-\Gamma & \omega_{m} & 0 & 0 \\
-\omega_{m} & -\Gamma & 2 g_{2} & 0 \\
0 & 0 & -\gamma & \tilde{\Delta} \\
-2 g_{2} & 0 & -\tilde{\Delta} & -\gamma \end{array} \right)
\end{equation}

The formal solution of Eqn.(17) is given by $R(t)= F(t) R(0)+ \int_{0}^{t} ds F(s) N(t-s)$ with $F(t)=e^{Mt}$. The system reaches a steady state only if it is stable, which is possible when all the eigenvalues of the drift matrix $M$ have negative real parts so that $F(\infty)=0$. Consequently, one gets the following equation for the steady state correlation matrix (CM):

\begin{equation}
M V +V M^{T}=-D ,
\end{equation}

where, $V=\int_{0}^{\infty} ds F(s) D F(s) ^{T}$ and $D$ is the diffusion matrix given as $D=Diag [\Gamma(2 n_{b}+1), \Gamma (2 n_{b}+1), \gamma, \gamma]$. In order to study the conditions under which the optical mode and the vibrational mode of the ion are entangled, we compute the logarithmic negativity $E_{N}$ which is for the CV case is

\begin{equation}
E_{N}=max[0,\ln 2 \mu^{-}],
\end{equation}

where, $\mu^{-}=\frac{1}{\sqrt{2}}[A-(A^{2}-4 det V)^{1/2}]^{1/2}$, with $A=\sum(V)$ and $\sum(V)=det(X)+det(Y)-2det(Z)$. Here we have used the $2 x 2$ block form of the CM as:

\begin{equation}
 V = \left( \begin{array}{cc}
X & Y  \\
Z^{T} & Y\end{array} \right)
\end{equation}

A Gaussian state is entangled only if $\mu^{-}<1/2$ or $4 det(V)< \sum (V)-1/4$ and it is equivalent to Simon's necessary and sufficient entanglement non-positive partial transpose criterion of the Gaussian states.

Fig.4 shows the plot of logarithmic negativity $E_{N}$ of the coupled ion-cavity system versus the normalized detuning ($\Delta/\gamma$). The stationary entanglement between a single driven optical cavity field mode and a mechanical mode of the ion via radiation pressure is clearly seen in the same parameter range where the NMS exists. Hence, entanglement and mechanical cooling exists together. A similar observation was also noted for a zigzag ion chain \citep{14} .This observation indicates a strong correlation between the two modes when a coherent energy exchange is taking place. The stationary entanglement between a single driven optical cavity field mode and a mechanical resonator via radiation pressure has already been analyzed in \citep{vitali1}.

Now, we discuss the optomechanical quantum nondemolition (QND) phonon and photon detection scheme.  For the optomechanical QND phonon detection, the cavity mode is pumped with a laser and the transmitted signal is measured using a photodetector. The transmitted signal from the cavity mode should be  filtered first before measuring using a  photodetector.  Moreover, in order to detect the photon number within its lifetime, it is also required that the measurement time should always be less than the inverse of the product of cavity decay rate and the photon number. Here, the photon number in the detection mode corresponding to different phonon states is to be studied as described in \citep{ludwig}. This detection mode photon number should follow the time evolution of the mechanical mode. In this way, the continuous monitoring of photon counts at the photodetector gives the QND measurement of the phonon number.

\section{Conclusions}
In conclusion, we have shown that an optomechanical system formed by a cavity mode and the mechanical mode of an ion can coherently exchange energy and produce stationary entanglements in an experimentally accessible parameter regime. The steady state photon number in the presence of large laser pump intensity shows bistable behaviour and exhibits a dependence on the relative position of the CM of the ion and the standing wave. Bitability is absent when the ion is centered at the anti-node of the standing wave. The coupling of the mechanical oscillator, the cavity field fluctuations leads to the splitting of the normal mode into two modes (Normal Mode Splitting) indicating energy exchange between the two modes. Such a strongly coupled two-mode system showing steady-state entanglement can be exploited for the realization of quantum memories and quantum interfaces within quantum-communication networks.

\begin{acknowledgments}
A. Bhattacherjee acknowledges financial support from the University Grants Commission, New Delhi under the UGC-Faculty Recharge Programme.
\end{acknowledgments}

\end{document}